# Measuring Compliance with the California Consumer Privacy Act Over Space and Time

Van Tran
University of Chicago
Chicago, IL, USA
tranv@uchicago.edu

Aarushi Mehrotra
University of Chicago
Chicago, IL, USA
aarushi.mehrotra0@gmail.com

Marshini Chetty
University of Chicago
Chicago, IL, USA
marshini@uchicago.edu

Nick Feamster
University of Chicago
Chicago, IL, USA
feamster@uchicago.edu

Jens Frankenreiter
Washington University in St. Louis
St. Louis, MO, USA
fjens@wustl.edu

Lior Strahilevitz
University of Chicago
Chicago, IL, USA
lior@uchicago.edu

## ABSTRACT

The widespread sharing of consumers' personal information with third parties raises significant privacy concerns. The California Consumer Privacy Act (CCPA) mandates that online businesses offer consumers the option to opt out of the sale and sharing of personal information. Our study automatically tracks the presence of the opt-out link longitudinally across multiple states after the California Privacy Rights Act (CPRA) went into effect. We categorize websites based on whether they are subject to CCPA and investigate cases of potential non-compliance. We find a number of websites that implement the opt-out link early and across all examined states but also find a significant number of CCPA-subject websites that fail to offer any opt-out methods even when CCPA is in effect. Our findings can shed light on how websites are reacting to the CCPA and identify potential gaps in compliance and opt-out method designs that hinder consumers from exercising CCPA opt-out rights.

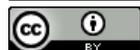

## 1 INTRODUCTION

In the past few years, states in the United States have begun enacting comprehensive consumer privacy laws modeled loosely on Europe's General Data Protection Regulation (GDPR) [16, 22, 55, 58]. The most ambitious of these state frameworks, the California Consumer Privacy Act (CCPA), was adopted in California. This act grants essential rights to consumers in that state, while placing obligations on larger for-profit companies that operate in the state and meet specific thresholds, such as annual revenues exceeding 25 million dollars or buying, selling, or sharing the personal data of 100,000 or more California residents [2, 15, 38].

The basic structure of the CCPA places the onus on consumers themselves to safeguard their privacy. Consumers can limit the sale and sharing of their personal information, find out what information companies possess about them, request the deletion of such information, or rectify errors concerning their personal information. Yet, to exercise these important CCPA rights, individual consumers need to affirmatively invoke them [2, 15, 17]. To facilitate consumers' ability to exercise their privacy rights, the CCPA also imposes disclosure obligations on businesses. For example, websites subject to the CCPA must typically create a link on their home pages that a consumer can click on to invoke their CCPA rights [5, 17].

In this paper, we evaluate the effectiveness of this policy design choice by focusing on the first step in that process—specifically, whether for-profit businesses comply with the CCPA's mandates to enable consumers to opt out of the sale and sharing of their personal information. If companies ignore their legal obligations to facilitate (and ultimately honor) opt-out requests, then the far-reaching privacy rights that the CCPA grants to consumers become illusory.

The CCPA's rules have changed over time, due to the enactment of statutory amendments (for example, via the California Privacy Rights Act (CPRA)) and regulations promulgated by California's Attorney General and Privacy Protection Agency, altering the requirements for opt-out methods (please refer to Figure 1 for important events of the CCPA and Section 2 for more details about how the CCPA evolved over time). It is important to monitor whether businesses subjected to the CCPA are staying current with the CCPA's shifting requirements. It is also valuable to assess whether companies extend these CCPA rights to consumers who are located outside California (i.e., spillover effects of the CCPA to

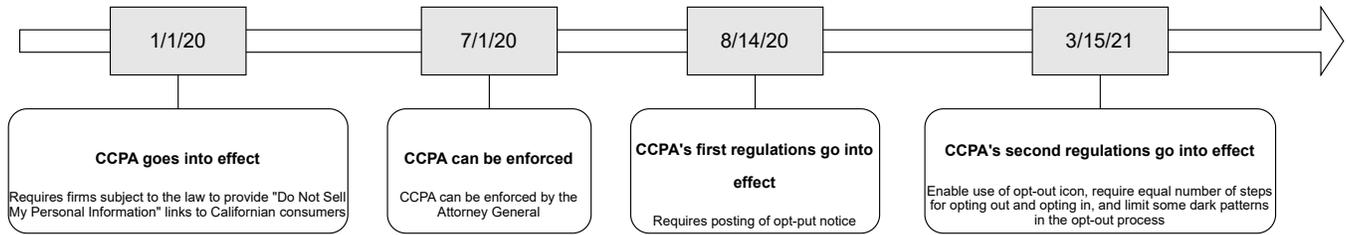

(a) Timeline of important events of the CCPA before Jan 1, 2023.

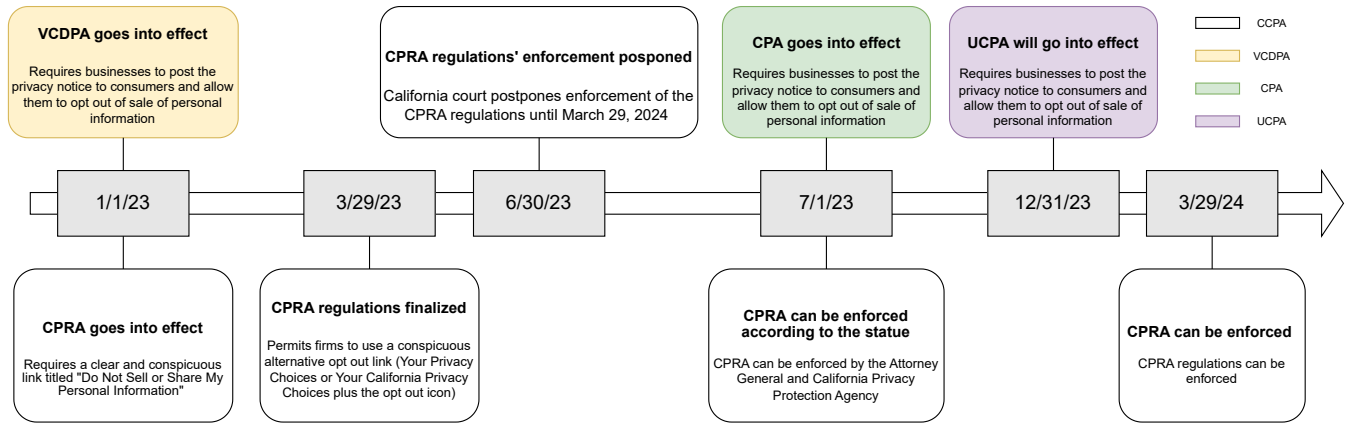

(b) Timeline of important events of the CCPA, the CPA, the VCDPA, and the UCPA, starting from Jan 1, 2023.

Figure 1: Timeline of important events for the California Consumer Privacy Act (CCPA), the Colorado Privacy Act (CPA), the Virginia Consumer Data Protection Act (VCDPA), and the Utah Consumer Privacy Act (UCPA) with the focus on events happening after Jan 1, 2023.

non-California consumers) and whether businesses not under CCPA jurisdiction offer CCPA rights to their consumers (i.e., spillover effects of the CCPA to non-CCPA-subject businesses). Note that measuring spillover effects to consumers in other states requires multiple steps, including answering the following questions: (i) do firms implement the opt-out links in states other than California? (ii) do they allow non-Californian consumers located in California to successfully opt out? and (iii) do they allow non-Californian consumers located in states other than California to successfully opt out? Given that steps (ii) and (iii) are difficult to measure, primarily because they require a determination of whether consumers can complete all the steps necessary to opt out successfully, this study only focus on measuring (i) only. To measure the spillover effects to non-CCPA-subject businesses, we focus on determining whether businesses that are probably not subject to CCPA implement the opt-out link for California consumers. Since several states enacted consumer privacy laws after California did, typically with less specific and onerous requirements, we are also interested in measuring whether businesses showed consumers in those states opt-out links mirroring those shown to Californians, or whether they tried to comply with other states' mandates in a narrower way. This study represents the first attempt to gauge businesses' compliance with the evolving CCPA requirements across space and time. Our research questions are as follows:

(1) How has the implementation of the CCPA's opt-out requirements changed over time?
(2) Do the opt-out links implemented on websites meet the CCPA's respective current requirements?
(3) Do the businesses that implement the opt-out link for California consumers also implement an opt-out link for non-California consumers?
(4) How does the implementation of the opt-out link differ between businesses that are subject to CCPA and those that are not subject to CCPA?
(5) What are the reasons for the absence of the opt-out link in websites that are likely subject to CCPA?

To answer these research questions, we designed and implemented a custom web measurement tool to survey 1,016 websites twice per month, from January-July 2023 from five vantage points: California (CA), Virginia (VA), Colorado (CO), Illinois (IL), and Utah (UT). We chose these vantage points because Virginia had a relatively lax law that was in effect for all of 2023, Colorado had a law almost as strict as California's that went into effect in mid-2023, Utah had a relatively lax law that went into effect at the very end of 2023, and Illinois had no data protection law remotely comparable to CCPA throughout our study period. We also rely on various data sources to examine whether the websites in our sample are likely subject to CCPA. Finally, we apply a combination of automatic and manual methods to check whether these websites implement alternative opt-out methods, as well as to account for the absence of opt-out links in websites that are likely subject to CCPA. Our findings are as follows:

- The number of websites starting to implement the CCPA opt-out link, as well as those changing their opt-out link to be CCPA-compliant, increases over time. By the end of July 2023, 349 (70%) out of 496 of websites likely subject to CCPA had implemented an opt-out link.
- Among websites that are likely subject to the CCPA but lack the opt-out link, less than 30% claim that they do not sell consumers' personal information while the rest either implemented the opt-out link later, embedded the opt-out methods in the privacy policy, offered offline opt-out methods, or did not mention CCPA or the opt-out rights of consumers at all.
- Among websites that implement the opt-out link regardless of whether they are subject to CCPA, more than 40% of them do not meet the updated CCPA's requirements for the appearance of the opt-out link in terms of terminology, locations, and other requirements such as including the opt-out icon when needed.
- There is also a varying degree of spillover effects on non-California consumers. Of the 581 websites that implement the opt-out link in at least one state by the end of our measurement period, the majority implement the opt-out link across all states, including Illinois (IL), which does not have a state-specific privacy law. On the other hand, some websites implement the opt-out link exclusively in a limited number of states.
- A significant number of for-profit websites not likely subject to the CCPA still implement the opt-out link.

| Online Opt-out Method | Acceptable Wording(s) | Other Requirements |
|---|---|---|
| "Do Not Sell or Share My Personal Information" opt-out link | "Do Not Sell or Share My Personal Information" | Link should be conspicuous and located at header or footer. |
| Alternative opt-out link | "Your Privacy Choices" or "Your California Choices" | Opt-out icon next to the link, link should be conspicuous and located at header or footer. |
| Frictionless opt-out preference signals | Not Applicable | Mention frictionless opt-out preference signals in privacy policy. |

Table 1: CCPA-compliant online opt-out methods. Online businesses are required to implement at least one of these methods.

This paper makes the following contributions:

- We design and implement a software tool to scalably, reliably, and efficiently measure the presence (or absence) of opt-out links for the sale or sharing of personal data on online websites.
- We use this tool to measure how businesses' implementation of CCPA's requirements change over time from multiple vantage points.
- We measure the extent of "spill-over" effects of CCPA compliance around the sale and sharing of personal data to non-California consumers and businesses not likely subject to CCPA, as well as jurisdictions that have data protection laws with less specific opt-out requirements than California.
- We present the reasons for the absence of the opt-out link for websites that are likely subject to CCPA but do not implement the opt-out link.

Our findings suggest that the California Consumer Privacy Act (CCPA) impacts not only businesses directly subject to it, but also those that are not. The study also reveals positive spill-over effects in states with less stringent privacy laws, and even in those without state specific privacy regulations. Additionally, our study provides clear evidence of non-compliance among some websites, underscoring the need for strict CCPA enforcement to ensure adherence. The rest of the paper is organized as follows. Section 2 presents background and related works. Section 3 describes our data collection process, including our rationale for the selection of the websites that we chose to study. Section 4 describes our measurement methods for studying the presence of opt-out links over time, as well as the presence of alternative opt-out methods. Section 5 presents our findings, Section 6 discusses the implications of our findings, and Section 7 concludes.

## 2 BACKGROUND AND RELATED WORK

In this section, we present background on the CCPA, as well as previous related works from both the computer science and legal research communities concerning privacy laws and compliance.

### 2.1 The California Consumer Privacy Act

Compared to other jurisdictions, especially the European Union, the United States has long been reluctant to enact omnibus consumer data protection laws [26]. While the European Union has robust regulations that afford consumers rights concerning a variety of organizations and businesses that manage their data [11], the U.S. has primarily relied on sector-specific laws. Examples include the Health Insurance Portability and Accountability Act (HIPAA) for healthcare and the Family Educational Rights and Privacy Act (FERPA) for student education records [24, 54]. This piecemeal strategy has been criticized for leading to potential gaps in protection

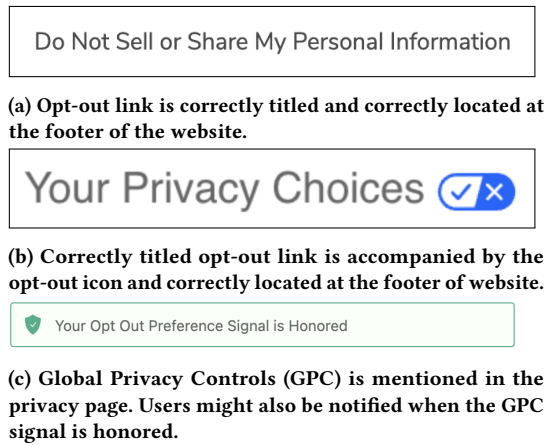

(a) Opt-out link is correctly titled and correctly located at the footer of the website.

(b) Correctly titled opt-out link is accompanied by the opt-out icon and correctly located at the footer of website.

(c) Global Privacy Controls (GPC) is mentioned in the privacy page. Users might also be notified when the GPC signal is honored.

Figure 2: Examples of CCPA-compliant online opt-out methods.

and failing to offer a holistic safeguard for consumers across all sectors, although defenders of the American approach note that US privacy laws historically have been enforced more vigorously than the EU's provisions [50, 51].

Although comprehensive data protection legislation at the federal level remains elusive, several states have taken initiative by enacting omnibus privacy laws. California blazed the trail with the introduction of the CCPA in 2018. The genesis of CCPA lies in a ballot proposition driven by Californian citizens. Its eventual realization was the outcome of a compromise: state legislators and the proposition's initiators agreed to the enactment of a slightly less stringent privacy law in lieu of the original proposition [6]. The CCPA became effective on January 1, 2020. Later in the same year, Californian voters approved another ballot proposition, the California Pivacy Rights Act (CPRA), which strengthened the protections of the original CCPA, introduced new measures, and created a specialized enforcement body, the California Privacy Protection Agency (CPPA) [32].

The CCPA's centerpiece is a set of "privacy rights" [55, 58]. Consumers are entitled to demand from firms that they are informed about the data collected about them and that such data is deleted on request. The CCPA also furnishes consumers with a right to ask that a firm stop "selling" (and, since the entry into force of the CPRA, "sharing") their information [38]. In this paper, we focus on the requirement that firms include a link on their home page that allows users to exercise this latter right easily.

**Evolution of CCPA Over Time:** The CCPA took effect on January 1, 2020, requiring businesses that sell consumer personal data to offer clear opt-out options, including an accessible form via a "Do Not Sell My Personal Information" link on their websites or apps [31]. Applicable businesses are those for-profit entities conducting business in California that handle personal data and meet at least one of these criteria: (a) at least $25 million in annual revenue, (b) handle personal data of 50,000 or more California consumers, households, or devices annually for commercial purposes, or (c) earn at least 50% of their annual revenue from selling Californians' personal data [31].

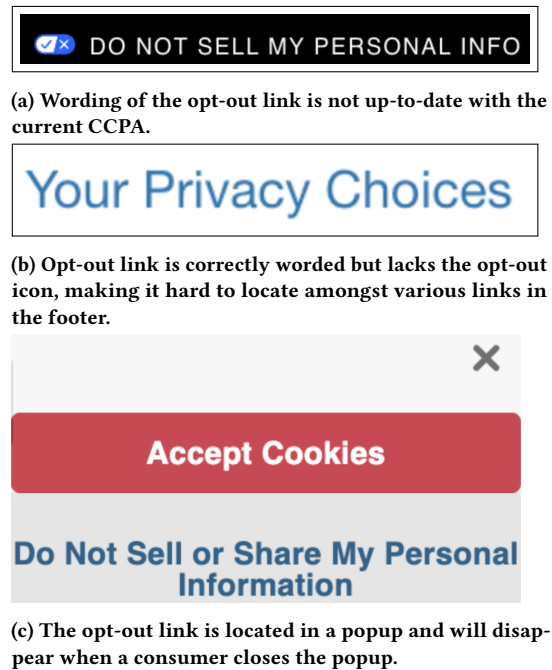

(a) Wording of the opt-out link is not up-to-date with the current CCPA.

(b) Opt-out link is correctly worded but lacks the opt-out icon, making it hard to locate amongst various links in the footer.

(c) The opt-out link is located in a popup and will disappear when a consumer closes the popup.

Figure 3: Examples of non-CCPA-compliant online opt-out methods.

The CPRA amends the CCPA and took effect on January 1, 2023. It revised the criteria, raising the threshold to 100,000 consumers, households, or devices, and expanded the mandate to businesses that share, not just sell, personal data, requiring them to add a "Do Not Sell or Share My Personal Information" link [15, 38]. It also provided an alternative, frictionless, browser-based preference signal (e.g. Global Privacy Control or GPC) that websites could implement instead of implementing the opt-out link [20, 53, 60]. Furthermore, the CPRA differentiates sensitive personal information from general personal data, requiring businesses that collect and process sensitive information to post an additional link titled "Limit the Use of My Sensitive Personal Information". Businesses can instead offer a consolidated "Privacy Choices" link in lieu of posting the two links mentioned above [18, 38, 53]. Table 1 details specific requirements for compliant online opt-out methods according the latest CCPA, updated on July 1, 2023.

The CCPA is explicit about the wording and appearance of the opt-out link and examples of opt-out links that fully meet these requirements are shown in Figure 2. In contrast, the examples such as those in Figure 3 appear to not be CCPA-compliant. Unlike the opt-out link where consumers click on the link on the website's homepage and follow steps such as toggling a button or completing a form to submit the opt-out request, in order to send a GPC signal, consumers need to utilize the privacy features on web browsers to send GPC signals to the website they are visiting. Depending on which browser they use, users may be required to do different steps to invoke GPC signals. For example, if users are browsing with DuckDuckGo [35] or Brave [30], GPC signals will be sent automatically to the website that they are visiting. If users are browsing with Google Chrome, they will need to install a browser extension before sending GPC signals [20, 53]. Businesses that honor the frictionless opt-out preference signals need to fully implement the opt-out process without changing the consumers' experience.

## 2.2 Other States' Privacy Laws

Following California's pioneering efforts with the CCPA, several other states have adopted laws that echo the CCPA's provisions, albeit with variations [1, 34, 39, 41, 52, 56]. The first states to follow suit were Colorado and Virginia, which in 2021 adopted the Colorado Privacy Act (CPA) and the Virginia Consumer Data Protection Act (VCDPA). The VCDPA went into effect on January 1, 2023, while the CPA went into effect in July 1, 2023. Connecticut and Utah adopted privacy legislation in 2022. The Utah Consumer Privacy Act (UCPA) goes into effect on December 31, 2023. As of September 2023, Delaware, Indiana, Iowa, Montana, Oregon, Tennessee, and Texas have also enacted new privacy laws, with many other states considering privacy legislation in their respective legislative chambers.

Although there are similarities between privacy laws of VA, UT, and CO, and the CCPA, there are also notable differences. In terms of who the law applies to, the CCPA only applies to for-profit companies but the CPA applies to non-profits as well [33, 34]. CO also does not impose a threshold on the gross revenue of the business or how much of the revenue comes from the sale of personal data [34]. VA also does not impose a threshold on the gross revenue of companies but sets a threshold on how much revenue comes from the sale of data [41]. UT, on the other hand, sets a gross revenue threshold of at least $25 million as a necessary condition while also requiring that an entity meet one of two thresholds for number of consumers whose data is processed, depending on whether the company derives most of its revenue from sales of consumer data.

In terms of what privacy rights consumers have, VA, CO, and UT only mention that consumers have the right to opt out of processing of personal data for targeted advertising, sale of personal data, or certain types of profiling [34, 39, 41]. As such, the CCPA seems to be more restrictive since it allows users to opt out of sharing of personal data as well. Unlike these states, CA also treats sensitive data as a special category of personal data, hence, allowing users to limit the use of sensitive data.

In terms of how consumers can exercise the right to opt out, CA is more explicit about opt-out methods that businesses have to implement such as the opt-out link, or the privacy preferences signals (GPC) [33]. California has spelled out precisely what "magic words" a website is to use if it does not employ a frictionless mechanism such as GPC. By contrast, VA, CO, and UT only require businesses to implement the privacy notice and allow users to submit opt-out requests, without explicit requirements about what language should appear on websites to satisfy these requirements. For a more comprehensive comparison of state specific privacy laws, please refer to Table 7 in the Appendix.

## 2.3 Research on User Privacy and Regulations

*2.3.1 Privacy-related Regulations.* There has been a growing body of literature in CHI and related communities measuring responses of entities to privacy-related regulations such as the Gramm-Leach-Bliley Act [10] and European regulations such General Data Privacy Regulation (GDPR) [12, 44]. For example, Nguyen et al. studies whether consent for processing of users' personal data is given freely and in an informed manner by automating the interaction with and collection of consent notices and analyzing those notices. Degeling et al. [12] measures the impacts of the GDPR on web privacy by monitoring and analyzing changes in the privacy policies of websites in European countries over time. Some work, on the other hand, examines how privacy-related regulations affect the user experience on different services and apps, and also whether different apps or services are compliant with regulations such as the Children's Online Privacy Protection Act (COPPA) [4, 21, 49, 59]. Some of these studies [45] also focus on the end-user's experiences with regulation-related interfaces such as consent notices and privacy policies.

*2.3.2 CCPA Regulations.* There is also an emerging literature studying different aspects of the CCPA regulations. Some work studies the process of opting out of the sale and sharing of personal information by recruiting participants to take part in the opt-out process and tracking self reports [43] or observing of how users go through the opt-out process [46]. This research highlights the burdens consumers face when going through the opt-out process and how these factors impact users' willingness to opt out. Other works focus on the design choices instead. Habib et al. [19] examines what link text and icons can most effectively communicate the presence of privacy-related information to consumers through a series of user studies. Siebel et al. [53], on the other hand, studies what design choices in the opt-out mechanisms will create high consumer satisfaction, which increases opt-out of sale rates. There is additional work [3, 8] studying the presentation and clarity of privacy disclosures.

*2.3.3 Measuring CCPA Compliance.* Most related to our work are studies that also measure the CCPA's opt-out methods. For example, Zimmeck et al. [60] measures the usability and enforceability of using GPC signals. Using a combination of a web crawler and a browser extension, they measure the presence of the opt-out link and responses of websites to GPC signals. Their study discovers a large number of websites that fail to respect GPC signals or fail to implement the opt-out link. Van Nortwick et al. [57] measures the presence of the opt-out link on 1 million websites. By performing parallel measurement from California and Massachusetts, the study discovers websites that use geofences to dynamically hide opt-out links for non-California residents. In addition, the study tries to examine whether websites are subject to the CCPA using various metrics such as the number of unique website visitors. Charatan et al. [7] hypothesizes that some textual revisions of CPRA potentially have negative impacts on CCPA opt-out rights and examines this hypothesis by measuring changes in websites' implementation of opt-out link and GPC signals during the period that the CPRA went into effect and how that affect consumers' awareness and understanding of businesses' data practices. They report finding a number of websites that stopped providing any opt-out mechanisms after the CPRA went into effect and an increase in the number of websites that only respect GPC signals. Our work differs from the previous research in several important ways:

(1) Our work extends the prior literature by covering the measurement of the opt-out link from January 2023 to August 2023 when CPRA went into effect. The timeliness of the measurement allows us to compare the behaviors of websites before and after each milestone event in the enactment of the CPRA.
(2) We conduct our measurements twice per month. This fine granularity allows us to observe the responsiveness of websites to each timeline of the CCPA and CPRA regulations. Taking numerous measurement of websites over time also allows us to observe the patterns in the behaviors of websites that a one-time measurement cannot do, and it allows us to track compliance with changing legislative and regulatory mandates.
(3) Our work measures the presence of the opt-out links across time and space. This allows us to observe the spillover effects of CCPA to non-CA states and compare whether there are any differences in the implementation of the opt-out links across different states, including jurisdictions with consumer privacy laws, those with enacted consumer privacy laws that have not yet gone into effect, and those with no comprehensive consumer privacy laws at all.
(4) To account for potential non-compliance of websites, we integrate (i) the measurement of the opt-out link, the measurement of the opt-out preference signals (GPC), (ii) checking the privacy policy for privacy opt-out related information, and (iii) integrating businesses' information to infer whether entities are likely subject to CCPA.

This paper complements a growing body of work that discusses the CCPA's implications and evaluates its effectiveness as a safeguard of consumer rights[27, 47, 48, 57].

## 3 DATASET

To answer our research questions, we created a dataset of websites on which we measured CCPA compliance, specifically concerning allowing users to opt out of the sale and sharing of personal data. We describe how we create the dataset and perform our measurements.
**Dataset:** To ensure the diversity of websites, we purposely selected popular websites from 80 different categories from a widely-used web ranking of websites [37], which provides information such as the overview, category, and number of unique visitors for websites. This helps to verify that websites we visit are legitimate, have a large number of users, and therefore, are likely subject to the CCPA. We examined 1,016 unique websites in total, as this is a large enough dataset to give a good coverage to tell us more about the compliance (or non-compliance) of businesses, while also being manageable enough for us to be able to measure these sites' compliance status frequently.
**Vantage Points:** Due to the lack of access to vantage points located physically at different states to measure the opt-out links frequently, we choose to use a Virtual Private Network (VPN) which allows us to connect to a network in a different location and browse websites as if we are browsing from that location itself. We conduct our measurements twice per month, starting from January 2023 to end of July 2023, from 4 states, namely Illinois (IL), California (CA), Colorado (CO) and Virginia (VA). We added Utah (UT) as another vantage point beginning in February 2023. UT, CO, and VA are states that had a comprehensive consumer privacy law in effect by the end of 2023 [34, 39, 41], though CO's is somewhat weaker in terms of protecting consumer privacy, and VA and UT's are much weaker [52, 56]. These states also have their privacy laws going into effect at different times, giving us insights into how websites respond to each new legal mandate. IL has no state specific privacy law resembling the CCPA [36], hence it serves as a comparison for states that have a state specific privacy law.

## 4 METHODS

In this section, we first describe how we check whether a website is likely subject to CCPA. Next, we explain how we automatically measure the presence of the opt-out link, followed by how the implementation of the opt-out preference signals and other opt-out methods are conducted.

### 4.1 Checking whether Websites Are Subject to CCPA

As previously mentioned, not all websites are subject to the CCPA. Further details on who is subjected to the CCPA can be found in Section 2. Certain criteria are difficult to check. For example, we can get the number of unique website visitors for each site we are measuring, but that does not necessarily correspond to the unique number of consumers, and especially California consumers. As such, during the curation of our dataset, we only select websites that serve at least 100,000 unique website visitors, based on website traffic analysis reports [37]. We then use Pitchbook [40], a financial data and software company that maintains comprehensive data on the private and public markets, to obtain information about each website including (i) annual gross revenue of the company, (ii) where the company's headquarters are located, and (iii) the number of unique visitors to the company's website. Unfortunately, not all companies are in Pitchbook's database, as such, we use another credible database from ZoomInfo [42] to obtain information such as the company's gross annual revenue, headquarter locations, and whether the company is government-owned, non-profit, or for-profit. Based on the criteria of who is subjected to the CCPA, we divide the websites into the following five groups (refer to Table 4 for details about how many websites fall into each category):

(1) **Non-profit (Not Subject to CCPA)**. These sites are either non-profit websites or are owned by the government. This group is clearly not subject to the CCPA.
(2) **Non-CCPA-subject For-profit (Not Subject to CCPA)**. The sites are for-profit websites with annual gross revenue less than $25 million. It is rather unlikely that entities in this group are subject to the CCPA because they do not meet the annual gross revenue threshold. However, if they annually buy, receive, sell, or share personal information of over 100,000 California residents, or derive 50% or more of their annual revenue from selling or sharing personal information of California residents, they are still subject to the CCPA [38].
(3) **CCPA-subject For-profit (Subject to CCPA)**. These sites are for-profit websites with annual gross revenue of at least $25 million and headquarters located inside the US. This

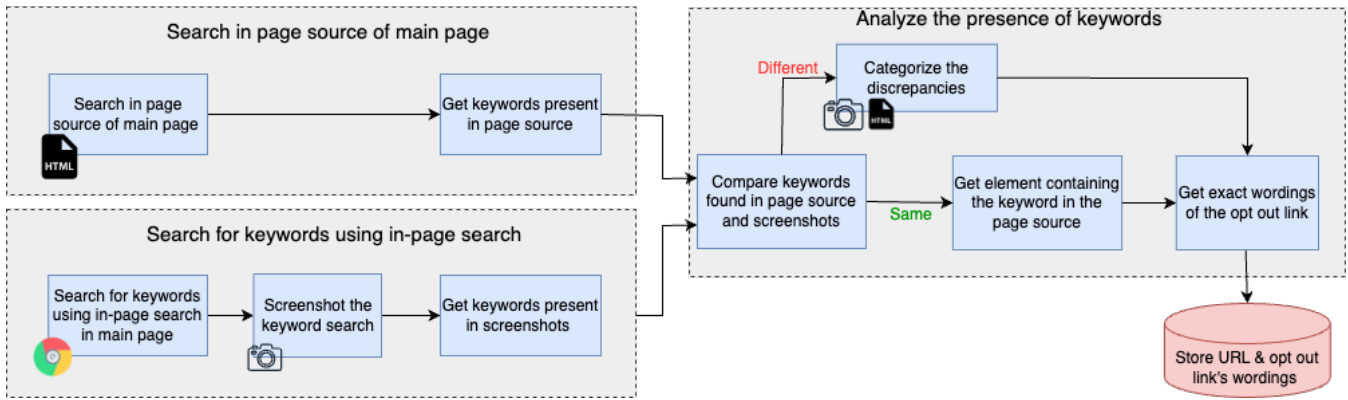

**Figure 4: Two-pronged approach to measure the presence and wording of the opt-out link.**

group is highly likely to be subject to the CCPA because their revenue exceeds the threshold and they are highly likely to conduct businesses in California.

(4) **International For-profit (Unsure whether They Are Subject to CCPA)**. These sites are for-profit websites with annual gross revenue greater than $25 million and headquarters located outside of the US. Assessing CCPA's applicability to this group is harder because their consumer bases may be international, so it is less certain whether they conduct businesses in California, how many California consumers they have, and how much of the total revenue comes from selling California consumers' personal information.

(5) **Unknown For-profit (Unsure whether They Are Subject to CCPA)**. These sites are for-profit websites whose information is not available, hence, it is difficult to conclude whether they are likely subject to CCPA.

## 4.2 Measuring the Presence of the Opt-out Link Over Time

This section first describes how the measurement of the opt-out link is conducted, followed by the performance of our measurement.

*4.2.1 Utilizing a Two-pronged Approach to Measure the Presence of the Opt-out Link:* Because the opt-out link's wording must follow the CCPA's requirements [38], we first constructed a list of compliant keywords based on the CCPA requirements. To help us construct this list, we manually visited a subset of the 1,016 websites and realized that some websites might use different wordings to refer to the opt-out link such as "Do not share", "Don't sell/share" or "for CA and VA Residents". We then conducted a more thorough manual inspection of all the websites in our dataset and added these additional phrasings to the original keyword search list. We use these phrasings to search for the presence of an opt-out link on the site. A detailed description of the method follows below.

Because we are measuring a large number of websites, frequently, over a long period of time, we must devise an efficient measurement method that enables retrieving exact phrasings of the opt-out link while providing screenshots to confirm the visual presence of opt-out link. We used a two-prong approach to confirm the presence of the opt-out link: one that involves direct searching of keywords on the web page and one that involves analyzing the page source of websites. Direct search on the website and page source analysis are not only fast but also complementary. Saving the page source allows us to find the exact wording used by the website. However, if the opt-out link is implemented in JavaScript, it is often the case that the page source cannot capture it. On the other hand, directly searching on the rendered website involves direct searching for opt-out keywords automatically using the 'ctrl'+'f' function on the browser. However, since it is time-consuming to perform direct searching on all websites, we can only do so for a limited number of keywords.

In addition to capturing the page source and directly searching on the rendered websites, we also capture screenshots of the websites during the direct search process to enable us to perform a manual check when the two approaches disagree. Our method to measure the presence of an opt-out link is described in Figure 4. After rendering each website using Selenium, we first check whether there is an error in accessing the webpage by searching for a few keywords indicating the presence of errors, including errors related to robot detection, network errors, or other blocking. Once the page is determined to be accessible, we conduct our search for the opt-out link as follows:

**Search in Page Source of the Main Page:** After the page is rendered, we save the page source of the main page. Initial inspection of these websites reveal that the opt-out link is often implemented as a <span>, <a> or <button> in the page source. As such, we search for any of these elements containing the keywords. We obtain the text of these elements, divide it into sentence or phrases and finally, narrow them down to the exact keywords of the opt-out link. If none of the keywords are present, we continue to search through all <div> to ensure that we do not miss out on any elements containing the keywords.

**Search for Keywords Using In-page Search:** We use the PyAutoGUI Python library to control the keyboard and mouse actions to conduct an in-page search for the keywords on the rendered page. For each keyword in the keyword list, we use the find in page function to search for the keyword on the rendered web page and take a screenshot of the web page for later verification. From the screenshot of each keyword search, we crop out the picture of the search bar and convert it to text using the PyTesseract library. We then search this text to see if the returned number of results that match with the keyword is greater than 0. If it is, we conclude that the keyword is present on the page.

**Analyze the Presence of Keywords:** If the keyword found in the in-page search process is also present in the page source, we conclude that the two methods agree. We store the website's URL and extract the opt-out link's wording from the page source. If they disagree, we do a manual check using the page source and screenshots of in page search to account for the discrepancy and then, get the exact phrasings of the opt-out link.

> "We do not sell OEM (Original Equipment Manufactorer) cartridges on this website."

(a) "Do not sell" is in the main page but it is not meant for sale of personal data. False positive case for website www.1ink.com.

> "This website uses cookies to personalize and improve your site experience, to enable us to analyze the traffic and usage of the website, to remember your settings and *privacy choices*, and to support our marketing and sales efforts."

(b) "Privacy choices" is in the main page but it is not used as an opt-out method. False positive case with website www.delonghi.com.

> "My Privacy and Ad Choices"

(c) Opt-out link is titled "My privacy and ad choices" which is slightly different from CCPA-compliant phrasings. False negative case with website www.venus.com.

> "Your California Privacy Rights"

(d) Opt-out link is titled "Your California Privacy Rights" which is slightly different from CCPA-compliant phrasings. False negative case with website www.intuit.com.

Figure 5: False positive and false negative cases of our method, measured from CA vantage point on July 8, 2023

*4.2.2 Examining the Performance of the Measurement Method:* To evaluate effectiveness of our measurement approach, we randomly select a date amongst the dates that we conduct our measurements. The date selected is July 8, 2023, conducted from a CA vantage point. After manually checking each website on that date, we compare the results of a manual check with those from our approach. Our approach yields two false positives (i.e., the opt-out link is not present but we report that it is present) and two false negatives (i.e., the opt-out link is present but we report that is it not present). Figure 5 shows more details about these cases. We also identify two cases where the opt-out link is present in the page source but is not present in the screenshot (for an example, website collegeboard.com has "Your Privacy Choices" present in the screenshot but not the page source) and 14 cases where the opt-out link is present in the screenshots but not the page source (refer to Figure 6 for an example). As such, this shows that our two-pronged approach can give higher confidence performance compared to using only page source or screenshot when checking the presence of the opt-out link.

### 4.3 Measuring Other Opt-out Methods

A business might not need to implement the opt-out link for several reasons, (i) it does not collect, sell, or share users' personal information, (ii) it processes the opt-out preference signals in a frictionless manner via a browser using GPC, or (iii) it does not collect users' personal information in an online manner and offers other offline opt-out mechanisms [46]. As such, we try to measure to what extent websites do not implement the opt-out link due to these reasons.

*4.3.1 Measuring the Presence and Notification of the Frictionless Opt-out Preference Signals.* In order to implement the frictionless opt-out preference signals that comply with the CCPA, businesses must (i) process the opt-out preference signals in a frictionless manner and (ii) mention in its privacy policy that it processes the opt-out preference signals in a frictionless manner. Businesses may also notify the consumers if the opt-out preference signals are honored [38]. To measure the presence and notification of the frictionless opt-out preference signals, we manually browsed each website from the CA vantage point on July 10, 2023 using the DuckDuckGo browser, which automatically sends the GPC signals to the website. We then visit pages where the frictionless opt-out preference signals are likely to be mentioned, including the front page, the opt-out page (if present, this page is obtained by clicking onto the opt-out link) and the privacy policy, and save each of these pages as a single file web page for further analysis. We then automatically extract the text of each page, filter out sentences that contain keywords related to the opt-out preference signals, and analyze these sentences to determine if the website mentions about the frictionless opt-out signals and whether they notify that the sent GPC signal is honored.

*4.3.2 Examining whether Businesses Sell/share Personal Information and whether They Offer Other Opt-out Methods.* We initially tried to scrape the privacy pages of these websites and use that data for this task but later realized that many websites such as www.apple.com place information related to the CCPA and California consumers in a separate page several steps away from the main Privacy page. Furthermore, the terminology of privacy policies can be nuanced and vary from website to website. As such, we did this process manually to ensure high confidence in our results. We only conduct our analysis on websites that are highly likely subject to the CCPA (for-profit websites whose headquarters are located within the US and whose revenue exceeds $25 million) but did not implement the opt-out link by the end of our measurement period. If businesses do not implement the opt-out link because they do not sell or share consumers' information, they have to mention this in their privacy

Figure 6: Opt-out link is present in the page source but is not seen in screenshots because it belongs to a div whose display style is set to None. Screenshot taken when inspecting the page source of https://usa.kaspersky.com/.

policy. As such, we visit each of these websites' privacy policy page, aiming to answer the following questions:

- Does the business sell/share consumers' personal information?
- Are any of the CCPA, opt-out rights mentioned in the privacy policy?
- Are there any methods for consumers to exercise the opt-out rights?

### 4.4 Limitations

**Measuring Compliance:** Our methods of checking whether websites are subject to the CCPA do not cover all criteria of who is subject to the CCPA. For example, we cannot capture whether businesses conduct business in California (although we believe that large companies, especially those with headquarters located in the US, likely conduct businesses in California), how many California consumers the company has, or how much of the entity's revenue derives from selling personal information. This information is not readily available.

**Checking if the Opt-out Link is Functional and Satisfies CCPA's Requirements:** Although our study measures compliance of websites with the CCPA, we do not examine whether the opt-out link is conspicuous enough and works effectively from the user's perspective. We only measure the presence of the opt-out link and consider sites compliant if the link is present. Given that websites' opt-out processes can be diverse, involving multiple steps and human understanding of the opt-out language to complete the opt-out process, measuring compliance remains a challenging task and is a topic for future work.

**Selecting Dataset:** Furthermore, while we choose to cap our dataset at 1016 websites for easier monitoring of these websites over time, we are aware that this number is still modest compared to those that are subject to the CCPA. As such, our numerical results might be affected by the selection of websites and may not reflect fully the percentage of websites that are CCPA compliant/non-compliant.

> "At this time, additional requests are not available in your state of residence. We're advocating for a national law that creates a consistent approach for all states."

Figure 7: IL residents cannot opt out if they click on the opt-out link. Screenshot of www.walmart.com after a user clicks on the link titled "Your Privacy Choices" and selects "IL" as the state of residence. Browsed from IL on July, 10, 2023.

**Checking Spillover Effects on Non-CA States:** The presence of the opt-out links in non-Californian states does not mean that a resident outside of California can successfully out out. For some websites that have the opt-out link present, if the users select that they are non-Californian residents, the opt-out of sale or sharing of personal information may not apply to them. Please refer to Figure 7 to see an example of such websites. However, once again, this checking of spillover effects for non-Californian residents is challenging to measure because it is difficult to automate and requires completion of the whole opt-out process to determine whether that can be done successfully.

**Checking Spillover Effects on Non-CCPA-subject Websites:** Although the presence of the opt-out link on for-profit websites whose revenue is less than 25 million dollars might be due to the spillover effects of CCPA on non-CCPA-subject websites, this might not be conclusive because our measure of what firms are subject to CCPA is underinclusive. That said, we strongly suspect that triggering the 25 million dollar revenue threshold is the reason why most firms are subject to the CCPA. Because figures about firms' number of California users and the percentage of firm revenue derived from the sale of personal data are not publicly available, it is quite challenging for researchers and agencies charged with enforcing the CCPA to develop a perfect measure of its applicability.

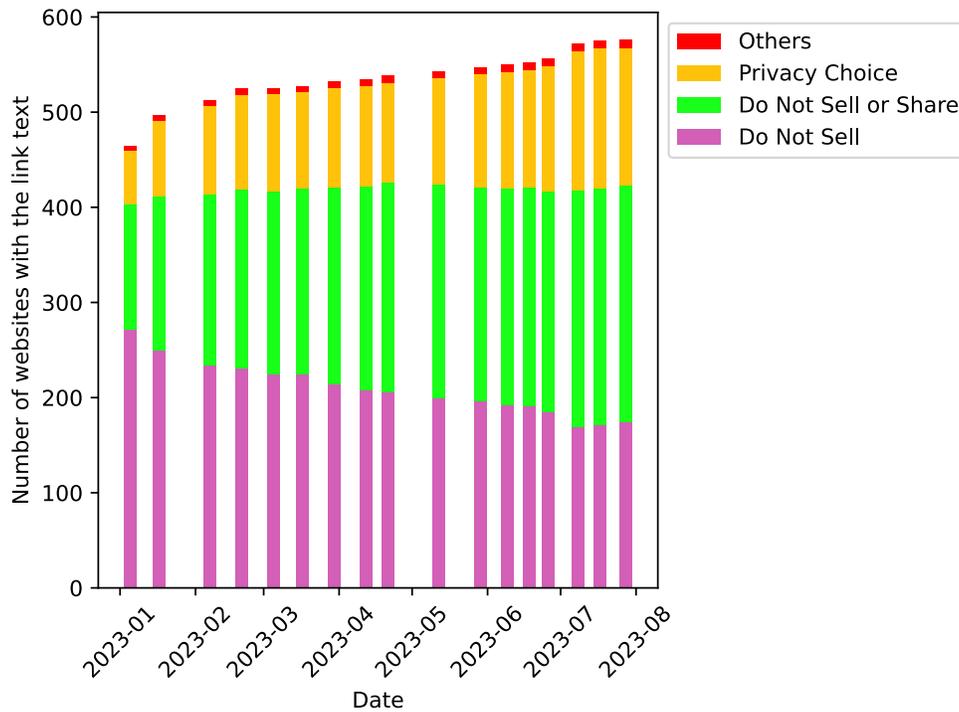

Figure 8: Number of websites where the opt out link is seen at least once by each measurement date, grouped by phrasings of the opt out link, measured from CA vantage point, N=1016.

## 5 FINDINGS

This section explores how the opt-out links for the sale and sharing of personal data are implemented over time, across different vantage points, and among the set of websites that we study. We explore several aspects: (i) how the presence of CCPA-compliant opt-out links changes over time; (ii) the extent to which the sites we study are (or are not) CCPA compliant; (iii) the extent to which the privacy regulations have resulted in "spillover effects" in other states that do not require opt-out links; (iv) the extent to which the privacy regulations have resulted in "spillover effects" in websites that are not likely subject to the CCPA; and (v) whether sites that do not implement the opt-out link offer alternative opt-out methods or are otherwise compliant.

Overall, we see an increase in the number of websites that have implemented the opt-out link for the first time or modify their opt-out link to meet the requirements of the CCPA. However, there are still a significant portion where the opt-out links have not met the CCPA's requirements or have not implemented any of the opt-out methods specified in the regulations. The rest of this section presents our findings in detail.

### 5.1 Implementation of the Opt-out Link Over Time for CA Vantage Point

Given the changes in the CCPA over time, we would like to see how the presence of the opt-out link for the sale and sharing of personal information varies over time from a CA vantage point. Figure 8 shows the number of websites where the opt-out link is seen at least once by each date of measurement. We see that by January 1, 2023, over 460 websites already have the opt-out link on their websites. This is understandable because CCPA went into effect on January 1, 2020. However, after CPRA went into effect in January 1, 2023, but before any government agency would enforce it, there was still an increase in the number of websites implementing the opt-out link. The increase is slightly steeper around January and February compared to other dates. We also observe that the phrasings of the opt-out links also change over time to meet the requirements of CPRA. We see that there is a decrease in the number of websites using the outdated wording "Do Not Sell My Personal Information", which satisfied CCPA requirements before it was amended by the CPRA, and there is an increase in the number of websites using CPRA-compliant wording: "Do Not Sell or Share My Personal Information" or "Privacy Choices". Refer to Table 2 for details about how the opt-out links are first seen and the most popular changes in the phrasings of the opt-out links. As such, overall, we see a tendency of websites to gradually implement the opt-out link as well as modify their opt out link to meet the up-to-date requirements of the CPRA.

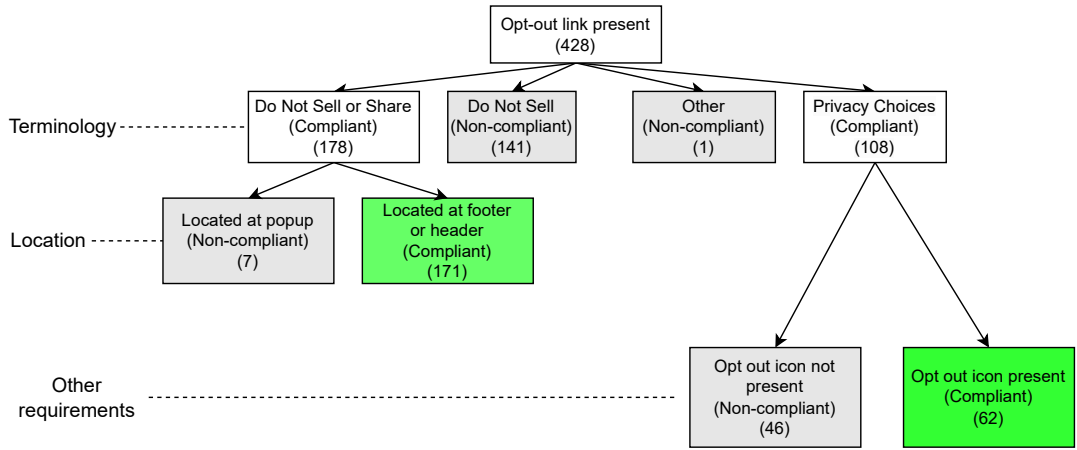

Figure 9: Compliance of opt out link's appearance to CCPA's requirements. Websites with the opt-out link are manually visited and checked on July 10, 2023 from CA vantage point. N=428.

Table 2: The most common phrasings of opt-out link when they first appeared (first 3 rows) and most common wording changes of the opt-out link (last 2 rows), observed from CA vantage point between January and July, 2023, N=576.

| Change in the opt-out link | # websites |
| --- | --- |
| No->Do Not Sell | 298 |
| No->Do Not Sell or Share | 169 |
| No->Privacy Choices | 101 |
| Do Not Sell->Do Not Sell or Share | 91 |
| Do Not Sell->Privacy Choices | 55 |

## 5.2 Compliance of the Opt-out Link with the Up-to-date CCPA's Requirements

The CCPA not only requires businesses to implement the opt-out link, but also to phrase and present the opt-out link in certain ways to help consumers easily locate it. To examine to what extent the implemented opt-out links fully meet CCPA's appearance requirements, we manually visited all websites that have the opt-out link from CA vantage point on July 10, 2023 and successfully visited 428 of them (We failed to visit the rest due to network errors, website maintenance, absence of opt-out links during measurement time, or because the website took too much time to load). (N = 572 websites implementing the opt-out link by July, 8, 2023). We manually checked for all requirements of the opt-out link, as detailed in Table 1. From Figure 9, we see that amongst these websites, 42% of them use the updated wording "Do not sell or share", 25% of them use the updated wording "Privacy Choices" and 33% still use the outdated wording "Do not sell". Amongst those that use the wording "Do not sell or share" (178/428), seven (4%) websites use popups to display the opt-out link, which might disappear when the user closes the window. It is unknown whether the website will show the popups again for the user. For the websites that use the wording "Privacy Choices" (108/428), 46 (over 40%) of them do not include the opt out icon, which might make it harder for users to locate the opt-out link. Overall, at the time of writing, even though CPRA already has been in effect for almost eight months, about 45% of websites that do implement the opt-out link still do not fully meet the requirements for the appearance of the opt-out link.

## 5.3 Variation of the Opt-out Links Across States and Spill-over Effects of CCPA

**Comparison of Implementation Trend Over Time.** Figure 10 shows that overall, the total number of websites with the opt-out link increases over time across all states, including IL. CA always has the largest number of websites with the opt-out link, followed by VA. This is possibly because the CCPA was introduced first. VA initially trailed behind CA by almost 100 websites but the gap between CA and VA gradually closed up by March, 2023. Interestingly, although the differences between CO, UT, and IL are not as significant when comparing them with CA or VA, there are changes in the order of which states have more number of websites implementing the opt-out link over time. As we move closer to July, 2023 where CO's CPA went into effect, we observe a pronounced increase in the number of websites implementing the opt-out link for CO, jumping CO ahead of UT and IL. This shows that there are many websites starting to implement the opt-out link around the date when they are required to. This pattern of implementation might reflect e-commerce firms deploying engineering resources close to the compliance deadlines to bring their organizations' actions into accordance with legal requirements.

**Comparison of Locations of the Opt-out Link.** Looking closely at the distribution of states where the opt-out link is seen (Table 3) in at least one state by the end of our measurement period (July, 2023), we observe that overall, 453 websites (78%), out of 581 of websites that implement the opt-out link in at least one state, implement the opt-out link across all states, including IL which does not have its own privacy law. The rest only implement the opt-out link in specific states. For example, 55 websites (9.5%) implement the opt-out link in CA and VA only, 25 websites (4.3%) implement the opt-out link for CA, VA, and CO only (the states where the state-specific privacy law already went into effect). 13 websites (2.2%) implement in CA, VA, CO, and UT only (the states that have a state-specific privacy law) and 9 websites (1.5%) implement in CA only. By the end of our measurement period (July, 2023), CPA had already been in effect for almost a month and VCDPA was already in effect for almost 8 months. The fact that there are websites that implement the opt-out link only in CA or only in CA and VA might indicate either websites are delaying extending the opt-out link to other states, or they are purposely not implementing the opt-out link in these states.

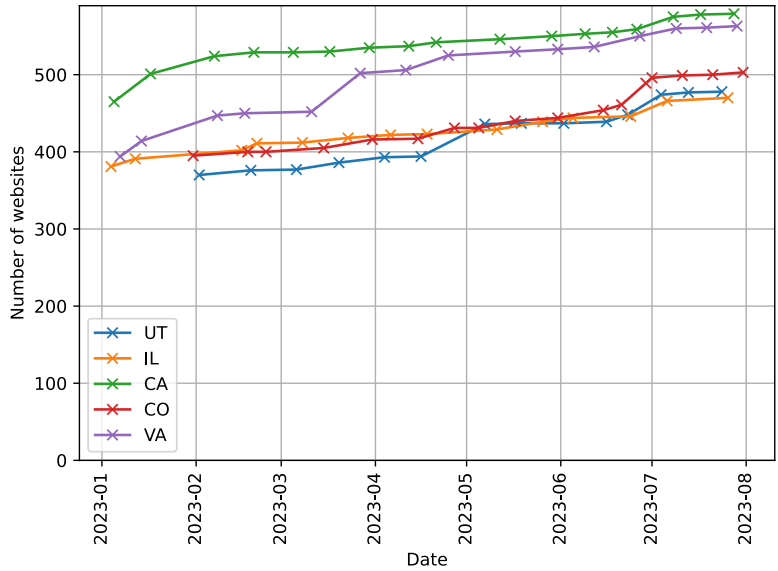

Figure 10: Number of websites where the opt-out link has been seen by each date at each state (N=1016 websites).

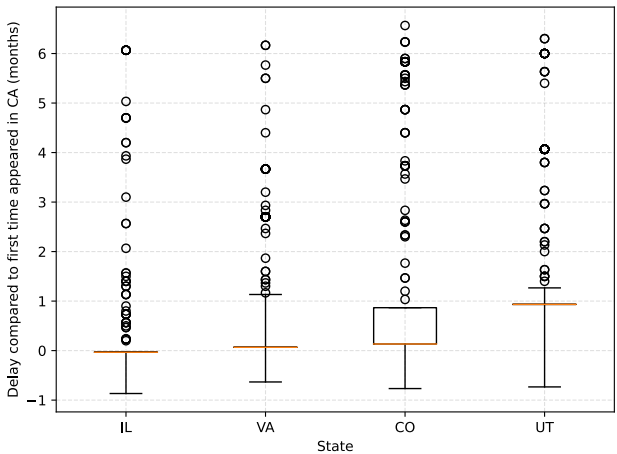

Figure 11: Whisker Box plot of delay (measured in months) of the implementation of the opt-out link in other states compared to CA, for the websites that implement the opt-out link in CA and at least 1 other state, whisker = [0–80]. Negative delay indicates that the opt-out link is observed earlier compared to CA. Delays of up to 1 month is most likely due to differences in measurement dates between different states.

Table 3: Distribution of states where the opt-out link is seen at least once during our measurement period, N=581.

| CA | VA | CO | UT | IL | # websites |
|----|----|----|----|----|------------|
| ✓ | ✓ | ✓ | ✓ | ✓ | 453 |
| ✓ | ✓ | ✓ | ✓ |   | 13 |
| ✓ | ✓ | ✓ |   | ✓ | 6 |
| ✓ | ✓ | ✓ |   |   | 25 |
| ✓ | ✓ |   | ✓ |   | 6 |
| ✓ | ✓ |   |   |   | 55 |
| ✓ |   |   |   |   | 9 |
| Other combinations of states |  |  |  |  | 14 |

**Comparison of Implementation Timings of the Opt-out Link.** For the websites that implement the opt-out link in more than one state, we compare the first date where the opt-out link is seen in other states versus when it is seen in CA (Refer to Figure 11). A general observation across all non-CA states is that while the majority of websites implement the opt-out link have small delays compared to CA (the whisker mostly lies between the range -1 and 1), there area a small number of websites that have large delays. Amongst these groups of websites with large delays, we observe that the largest delays seem to happen in CO, followed by VA, UT and lastly, IL. This could be because websites that do implement the opt-out link for all states, including IL, tend to implement it at the same time for all states. Websites that implement for specific states only tend to delay the implementation for the states that they have to. Hence, there are more websites that delay the implementation in VA by 1–2 months, in CO by around 6 months compared to the implementation in CA. Last but not least, because the UCPA had not gone into effect yet, websites did not have to implement the opt out link in UT. This results in a smaller number of websites with large delays.

Table 4: Number of websites in each category of websites and how many websites within each category have the opt out link ever seen during our measurement period (from Jan 2023 to end of July 2023) from CA vantage point.

| Group | | Description | Link present | Link absent | Total |
|---|---|---|---|---|---|
| Non-profit | | Government or non-profit | 2 | 64 | 66 |
| For-profit | | For-profit | | | |
| | Non-CCPA-subject | Revenue<$25 million | 190 | 170 | 360 |
| | CCPA-subject | **Revenue>=$25 million, headquarters in US** | **349** | **147** | **496** |
| | International | Revenue>=$25 million, headquarters outside US | 29 | 45 | 74 |
| | Unknown | Revenue is not available | 9 | 11 | 20 |

**Comparison of Phrasings of the Opt-out Link.** In terms of the phrasings of the opt-out link, most websites also use CCPA-compliant words for the opt-out link in non-California states except for few exceptions. For example, there are two websites (2/581) that use the wording "For CA and VA residents" and five websites (5/581) that use "opt out of sale" instead of CCPA-compliant phrasings for non-CA states. Interestingly, these five websites only implement the opt-out link for CA and VA, and sometimes CO. Briefly looking at the user interface of the website, we also see that there might be some slight differences after a user clicks on the opt-out link when browsing from different vantage points. An example of such interfaces can be found in Figure 12. As such, it tends to suggest that there are a small number of websites that implement differentiated opt-out links and offer differentiated privacy rights to their consumers, depending on where they reside.

### 5.4 Variation of the Opt-out Links Across Different Groups of Websites

We divide the websites into different categories. Table 4 shows the number of websites implementing the opt-out link for different categories. Interestingly, the website www.airforce.com (government website) and www.npr.org (non-profit website) implement the opt-out link although they are not required to. We also see that over half of the *non-CCPA-subject for-profit* websites implement the opt-out link. Although we do not have sufficient information to conclude whether these websites fall into other criteria of CCPA that we could not verify as explained in methods, this could also be an indication of CCPA's spillover effects on businesses that are not subjected to CCPA. Last but not least, 70% of the *CCPA-subject for-profit* implement the opt-out link. This category is highly likely subject to CCPA, unless they do not sell or share consumers' personal information. We will investigate those that fall into this category but fail to offer an opt-out link below.

Table 5: Number of websites that implement online opt-out mechanisms, manually measured on July, 10, 2023 from CA vantage point, N = 911. Most websites implement the GPC signals as a complement rather than replacement for the opt-out link.

| Opt-out link present | Use GPC signals | # websites |
|---|---|---|
| ✓ | ✓ | 205 |
| ✓ | | 223 |
| | ✓ | 34 |
| | | 449 |

### 5.5 Do Businesses that Do Not Implement the Opt-out Link Offer Alternative Opt-out Methods or Do Not Sell/share Personal Information?

We observe that although the effective date of the CCPA was January 1, 2023 (almost eight months before the end of our measurement period), almost 30% of websites that are highly likely subject to the CCPA still do not implement the opt-out link yet. This motivates us to investigate the reasons for this potentially high non-compliance rate.

*5.5.1 Frictionless Opt-out Preference Signal.* Table 1 shows that another acceptable online opt-out method for businesses to be CCPA-compliant is the frictionless browser-based opt-out preference signal such as Global Privacy Controls (GPC). We measured a snapshot of the implementation of the GPC on July, 10, 2023, near the end of our measurement period by manually browsing each website using a private browser, DuckDuckGo from CA vantage point. We managed to successfully access 911/1016 websites (our scraper failed to access some websites because of network errors, website maintenance, or simply taking too much time to load) and found out that almost a quarter of them (239/911) implement the GPC signals. It is important to note that websites tend to use GPC signals as a complement to the opt-out link rather than replacement. In fact, 205 websites (86%) implement both GPC signals and an opt-out link and only 34 (14%) implement the GPC signals alone without the opt-out link. Please refer to Table 5 for more detailed results.

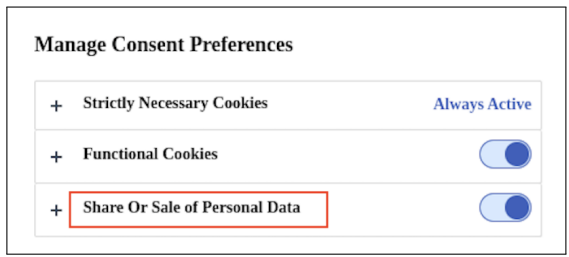

(a) Consumer can opt out of both selling and sharing of personal information if browsed from CA. Screenshot taken on July 29, 2023 from CA vantage point.

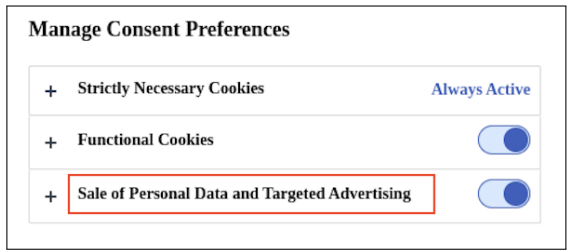

(b) Consumer can only opt out of targeted advertising and selling of personal information if browsed from VA. Screenshot taken on July 28, 2023 from VA vantage point.

Figure 12: Website offers differentiated versions of the opt-out choices for different states. Screenshots of website www.lumen.com after clicking opt out link from CA vantage point (left) and VA vantage point (right).

*5.5.2 For CCPA-subject For-profit Websites that Do Not Implement the Opt-out Link, Do They either Not Sell or Share Personal Information, Offer Other Opt-out Methods, or Are Otherwise Compliant?* Since it is hard to conclude with certainty who is subject to the CCPA, we focus on investigating the *CCPA-subject for-profit* websites that do not provide the opt-out link. This group of websites totals up to 148 websites (refer to Table 4). Note that we decide to do this manually because for some websites, we might need to perform several steps before reaching relevant information related to the CCPA. We encode the reason for why the opt-out-link is absent. Results can be found in Table 6. We found that only 43/148 of these websites do not implement the opt-out link because they indicate that they "do not sell" consumers' personal information. There are also 12 websites have the opt-out link on the checking date, indicating that they implement the opt-out link way later than when CPRA took effect. Interestingly, there are six websites that do mention the CCPA in their privacy policy but do not mention opt-out rights or opt-out methods. There are 37 websites that do offer some opt-out methods such as using a contact form, sending an email, making a phone call, or sending mail but users have to identify these opt-out methods by reading the privacy policy. Lastly, we cannot locate the privacy policy for five websites and sections related to CCPA or opt-out rights for 41 websites. Please refer to Figure 13 for interesting examples observed during our measurement process.

Table 6: Reasons for the absence of the opt-out link during our measurement period of websites that are for-profit, revenue exceeding $25 million US dollars and headquarters located in the US. Conducted on November 1, 2023 from CA vantage point, N=147.

| Reason | # websites |
| --- | --- |
| Does not sell personal information | 43 |
| No mention of CCPA or opt-out rights | 41 |
| Opt-out methods found in privacy policy | 37 |
| Late implementation of opt out link | 12 |
| Mention CCPA but no opt-out rights/methods | 6 |
| No privacy policy found | 5 |
| Error/blocked/website retired | 3 |
| Total | 147 |

## 6 DISCUSSION

We explored various implications of our findings, addressing challenges encountered in conducting this large-scale, automated study, and discussing the implications for regulators and others in automated compliance checking.

### 6.1 Consumer Protection Policy Implications

At the end of 2022, with Democrats about to lose their majority in the United States House of Representatives, Speaker Nancy Pelosi faced a choice. The House Energy and Commerce Committee voted by a 53–2 margin to approve the American Data Protection and Privacy Act (ADPPA) [23]. The ADPPA would have created sweeping new consumer privacy protections at the federal level. If enacted, the United States would have a new law, national in scope, to rival Europe's GDPR. But ADPPA would have preempted more protective state laws; this preemption provision was necessary in order to secure support from Congressional Republicans. This preemption provision meant that parts of CCPA might become invalid (because they were incompatible with the ADPPA), and California's ability to enact new laws in the future expanding consumer privacy rights further would be substantially curtailed. (Those laws would likely be preempted by ADPPA).

Faced with opposition from California's governor and chief privacy regulators [29]. Speaker Pelosi decided to block the ADPPA. As a result, no bipartisan comprehensive privacy law was enacted, and such legislation has languished subsequent to the Republican takeover of the House. Our research is a first step towards evaluating the wisdom of Pelosi's decision. If California privacy laws are de facto national privacy laws, because companies will find it economical to give consumers nationwide the benefits of rights that Californians enjoy, then Pelosi almost certainly made the right call. If CCPA has had minimal spillover effects, then Pelosi's decision might have benefited Californians' privacy rights at the expense of their fellow Americans' rights.

Our study reveals that the CCPA does have a positive effect on the behaviors of websites, as evident by the increase in the implementation of the opt-out link over time, not only in CA, but also

> *"The following list details some of the major privacy regulations recognized by Adorama...*
> *–California Consumer Privacy Act (CCPA)"*

**(a) Www.dorama.com mentions that it recognizes the CCPA but does not mention about CCPA opt-out rights or opt-out methods.**

> *"California consumers & CCPA:*
> *California consumers may make a request pursuant to their rights under the CCPA by contacting us at hello@ticketcity.com. We will verify your request using the information associated with your accounts, including email address. Government identification may be required."*

**(b) Www.ticketcity.com does not mention what rights consumer can exercise under the CCPA. Opt-out method is also unclear.**

> *"If you prefer not to receive interest-based advertising, you can opt-out by making a request via this **request form** or contacting us at **privacy@birchbox.com** or setting up a call with us, here."*

**(c) Www.birchbox.com has no opt-out link on the main page but has opt out methods located in privacy policy. Clicking the on the request form mentioned in privacy policy leads to a dysfunctional page.**

○ Details on how to opt out or learn more about your personal information can be found on our **Do Not Sell** My Personal Information page or email privacy@savage.ventures

## Your rights with respect to personal data

You have the following rights with respect to your personal data:

○ You may submit a request for access to the data we process about you
○ You may object to the processing
○ You may request an overview, in a commonly used format, of the data we process about you
○ You may request correction or deletion of the data if it is incorrect or not or no longer relevant, or to ask to restrict the processing of the data

**(d) Rare.us does not have an opt-out link on the main page. The privacy policy mentions that consumers can click on the "Do not Sell My Personal Information" page to learn about how to opt out. However, once clicking that page, no opt-out rights or opt-out methods are mentioned.**

**Figure 13: Examples of non-compliance to the CCPA.**

for non-CA states. However, the study also reveals a lag, and potentially non-compliance, in websites that fail to update their practices to align with evolving CCPA requirements. These findings for the CCPA mirror in some respects research on the General Data Protection Regulation (GDPR), a European Union privacy law for entities that conduct collect and process personal information of residents of the EU [9, 13, 25, 28]. For example, as a result of GDPR, the number of websites that have privacy policies increased over time, as did the number of websites with cookie consent banners [12]. However, much like how our study suggests partial compliance with the CCPA, there are still a number of websites not implementing the GDPR requirements. In addition, we also see a small number of websites offering delayed or differentiated versions of the opt-out links for residents in different states, just as websites offer differentiated versions of the privacy and cookies notices for residents in different countries in Europe [12]. This suggests that websites tend to have the same kind of behaviors in reacting to different privacy laws, at least with respect to laws like CCPA that entail moderate compliance costs. Another interesting observation during the course of our study is that, we observe some websites clearly acknowledge the CCPA, but fail to comply with the CCPA's requirements. Some examples include those that embed the link to opt-out forms in the privacy policy instead of using a conspicuous opt-out link on the main page, offer offline opt-out methods despite having an online platform, or merely offer instructions about how a user can modify various privacy settings. As such, these websites appear to violate CCPA's requirements despite knowing that they have to comply with the CCPA. These observations, perhaps, suggest that more muscular enforcement and significant fines are needed to spur these websites into updating their websites to be CCPA-compliant.

One recent study conducted during the same time period arrives at conclusions that differ from ours in some respects. While our study reveals an increase over time in the number of websites that implement the opt-out links or modify opt-out links to become CCPA-compliant (refer to Figure 8), Charatan et al. [7] discover a significant number of websites that stopped providing any opt-out methods after CPRA went into effect. They attribute this effect to CPRA's increase in the number of California consumers, households, or devices whose personal information must be collected for a website to be subject to the law's requirements (Section 4.3.1). (CCPA applied to websites that buy, sell, or share the personal information of 50,000 consumers, households, or devices, whereas CPRA raised that threshold to 100,000.) They also report that CPRA's modifications to the threshold caused websites to switch to respecting GPC signals only rather than providing opt-out links, reducing users' awareness and understanding of their privacy choices (Section 4.3.1). We hypothesize that the discrepancies result from these dynamics:

(1) Choice of Dataset. While our study focuses on a smaller number of websites that are likely subject to CCPA and CPRA because of the large scope of their online business, their study is performed on a much larger dataset of websites. As such, their study includes websites that are more likely to be affected by CPRA's changes to the reach of the statute. As the authors note, their set of 25,000 popular websites also includes sites not rendered in the English language and not aimed at consumers, suggesting that they are more likely to study websites that were subject to neither CCPA nor CPRA. At most, 15.2% of the websites in their sample provided a CCPA opt-out mechanism via a link or GPC. (Calculation based on Section 4.3.1).

(2) Missing Measures of CCPA Applicability. Relatedly, while our study uses independent measures of corporate revenues and for-profit status to exclude websites that are probably not subject to CCPA's requirements, their study does not have a robust method to check whether websites are subject to CCPA or are affected by the changes in the statutory scope. It is important to note that not all websites subject to CCPA comply with mandatory opt-out requirements, so when a website eliminates an opt-out process, that could indicate that it was subject to CCPA but not CPRA, or it could indicate noncompliance with the law. Ascribing significant changes in website behavior during the observation period to the changes in statutory scope is particularly tricky because the number of consumers, households, and devices whose personal information is collected by a firm is typically non-public information. Our effort to obtain the list of websites used in their study for comparison purposes was unsuccessful.

Both studies show that the number of websites that implement at least one opt-out method increased over time (refer to Figure 8 of our paper and Figure 2 of their study). Ultimately, despite our divergent interpretations of the data, the two studies' findings complement each other by providing varied perspectives into the impacts of CCPA and CPRA on different subsets of websites. The somewhat varied results also underscore the importance of examining the impacts of CCPA from multiple perspectives for a more comprehensive understanding of the impact of the laws.

### 6.2 Challenges in Compliance Checking

Confirming the presence or absence of a CCPA-prescribed opt-out link on websites is technically challenging. Determining compliance with the CCPA is even more complex, particularly for sites without an opt-out link. The difficulties include:

- Identifying firms subject to the CCPA is time-consuming and complex, especially when key information, like the number of California consumers or revenue from selling personal information, is not publicly available.
- Privacy policies are often lengthy, disorganized, and filled with confusing jargon. This makes it hard to determine whether websites are selling or sharing personal information. The updated CCPA complicates this further, as websites previously compliant might now fall short due to sharing consumer data for targeted marketing.
- Variations in state specific privacy laws add another layer of complexity. Websites might comply with some state laws but not others. This disparity makes it challenging to ascertain applicable laws, particularly when websites offer different privacy rights or opt-out links in various states.
- Measuring compliance with CCPA is easier than measuring compliance with other states' laws. Because California identifies two sets of "magic words" that will communicate opt-out rights to consumers, researchers and agencies charged with law enforcement can more readily automate the process of measuring compliance. By contrast, laws like Virginia's or Utah's tell entities to inform consumers about their rights, but give them lots of flexibility in what wording to employ. There does not seem to be any obvious benefit to permitting this flexibility. It makes it harder for consumers to find what they are looking for and harder for third parties to measure compliance.

These challenges in automated consumer privacy law compliance monitoring highlight the difficulties faced not only by consumers in invoking privacy rights but also by regulators in enforcing compliance efficiently. In the following section, we delve deeper into these policy implications and offer suggestions for addressing them.

### 6.3 Recommendations for Policymakers

The challenges in checking CCPA compliance suggest the need for more accessible compliance monitoring. Proposed measures include:

- Creating a public dataset listing entities subject to the CCPA to simplify monitoring compliance.
- Mandating a standardized structure for privacy policies, enabling quicker access to relevant sections.
- Standardizing the language used in privacy policies to require a finite list of "magic words", particularly when businesses do not sell or share personal information.

In addition, given these evident non-compliance observations and the time elapsed since the enactment of the CCPA and the CPRA, we strongly recommend that the California Attorney General's Office use automated techniques to identify entities that are non-compliant and demand documentation from these firms that CCPA is inapplicable to them.

## 6.4 Avenues for Future HCI and Policy Research

The CCPA places the responsibility of opting out on consumers, making it crucial for them to decide whether to exercise their privacy rights. This decision is influenced by factors such as consumer awareness about privacy rights and their preference for certain opt-out methods. Future research could extend this work to explore how aware users are of their privacy rights and their ability to navigate different opt-out processes. It could also compare users' familiarity with manual opt-out methods, like using an opt-out link, against automated techniques such as browser-based Global Privacy Control settings.

Our study serves as a basis for measuring the spillover effects of the CCPA and understanding how websites respond to privacy laws in different states. It provides some reason to believe that CCPA may function as a de facto national comprehensive consumer privacy law, despite its ostensibly limited jurisdictional reach. In further research, we plan to exercise CCPA opt-out rights on behalf of consumers based in California and in Illinois. This will allow us to confirm that the potential spillover effects we observe are real and meaningful. Our research indicates some websites restrict opt-out options for non-CA consumers or offer varied opt-out experiences, but the prevalence of these practices is unclear, and the presence of an opt-out link does not necessarily mean that opt-outs from non-Californians will be processed and respected.

At this juncture, though, there is some reason to believe that a "Sacramento effect" exists in consumer privacy law [14]. The California market is big enough, and providing different versions of web sites to residents of different states is cumbersome enough, to justify giving residents of every state the ability to exercise CCPA opt-out rights. If further explorations confirm this possibility, then it suggests that enacting federal legislation is a less urgent priority than many privacy advocates believe. In short, Nancy Pelosi's decision to kill the ADPPA may, in the long run, prove to be a decision that expands privacy protections for all Americans.

## 7 CONCLUSION

In this paper, we present an efficient and reliable method to monitor the presence of opt-out links for the sale and sharing of personal data on online businesses. We study compliance with the CCPA requirements of websites and monitor how compliance changes over time after this law went into effect. We discover that a significant portion of websites still fail to meet the CCPA requirements and investigate the reasons for that. We also discover variation in spillover effects, where a large portion of websites implement the opt out links across all states investigated, while some websites only provide the opt-out links in certain states. Our work builds up on previous work trying to measure websites' compliance to the CCPA by measuring the responses of websites during the period where CPRA went into effect. In addition, our study also offers new interesting insights into the extent of CCPA spillover effects to other states as well as websites that are likely not subject to CCPA. We provide some suggestive evidence that the CCPA may be de facto national consumer privacy legislation.

## REFERENCES

[1] Folks Andrew. 2023. US State Privacy Legislation Tracker. https://iapp.org/resources/article/us-state-privacy-legislation-tracker/. [Online; accessed 28-November-2023].

[2] Jeeyun Sophia Baik. 2020. Data Privacy Against Innovation or Against Discrimination?: The Case of the California Consumer Privacy Act (CCPA). *Telematics and Informatics* 52 (2020). https://doi.org/10.1016/j.tele.2020.101431

[3] Vinayshekhar Bannihatti Kumar, Roger Iyengar, Namita Nisal, Yuanyuan Feng, Hana Habib, Peter Story, Sushain Cherivirala, Margaret Hagan, Lorrie Cranor, Shomir Wilson, Florian Schaub, and Norman Sadeh. 2020. Finding a Choice in a Haystack: Automatic Extraction of Opt-Out Statements from Privacy Policy Text. In *Proceedings of The Web Conference 2020* (Taipei, Taiwan) *(WWW '20)*. Association for Computing Machinery, New York, NY, USA, 1943–1954. https://doi.org/10.1145/3366423.3380262

[4] Elijah Robert Bouma-Sims, Megan Li, Yanzi Lin, Adia Sakura-Lemessy, Alexandra Nisenoff, Ellie Young, Eleanor Birrell, Lorrie Faith Cranor, and Hana Habib. 2023. A US-UK Usability Evaluation of Consent Management Platform Cookie Consent Interface Design on Desktop and Mobile. In *Proceedings of the 2023 CHI Conference on Human Factors in Computing Systems* (<conf-loc>, <city>Hamburg</city>, <country>Germany</country>, </conf-loc>) *(CHI '23)*. Association for Computing Machinery, New York, NY, USA, Article 163, 36 pages. https://doi.org/10.1145/3544548.3580725

[5] Diane Y Byun. 2019. Privacy or Protection: The Catch-22 of the CCPA. *Loy. Consumer L. Rev.* 32 (2019), 246. https://heinonline.org/HOL/LandingPage?handle=hein.journals/lyclr32&div=13&id=&page=

[6] Anupam Chander, Margot E. Kaminski, and William McGeveran. 2017. Catalyzing Privacy Law. *Minnesota Law Review* 105 (2017), 1733–1802.

[7] Jan Charatan and Eleanor Birrell. 2023. Two Steps Forward and One Step Back: The Right to Opt-out of Sale under CPRA. arXiv:2312.15094 [cs.CY]

[8] Rex Chen, Fei Fang, Thomas Norton, Aleecia M. McDonald, and Norman Sadeh. 2021. Fighting the Fog: Evaluating the Clarity of Privacy Disclosures in the Age of CCPA. In *Proceedings of the 20th Workshop on Workshop on Privacy in the Electronic Society* (Virtual Event, Republic of Korea) *(WPES '21)*. Association for Computing Machinery, New York, NY, USA, 73–102. https://doi.org/10.1145/3463676.3485601

[9] Raffaele Congiu, Lorien Sabatino, and Geza Sapi. 2022. The Impact of Privacy Regulation on Web Traffic: Evidence From the GDPR. *Information Economics and Policy* 61 (2022), 101003.

[10] Lorrie Faith Cranor, Pedro Giovanni Leon, and Blase Ur. 2016. A Large-scale Evaluation of U.S. Financial Institutions' Standardized Privacy Notices. *ACM Transactions on the Web (TWEB)* 10, 3 (Aug 2016), Article 17, 1–33. https://doi.org/10.1145/2911988

[11] BØrge Dahl. 1993. Consumer Protection Within the European Union. *Journal of Consumer Policy* 16, 3-4 (1993), 345–353. https://doi.org/10.1007/BF01018761

[12] Martin Degeling, Christine Utz, Christopher Lentzsch, Henry Hosseini, Florian Schaub, and Thorsten Holz. 2018. We Value Your Privacy ... Now Take Some Cookies. *Informatik Spektrum* 42 (2018), 345 – 346. https://api.semanticscholar.org/CorpusID:52011984

[13] Martin Degeling, Christine Utz, Christopher Lentzsch, Henry Hosseini, Florian Schaub, and Thorsten Holz. 2018. We Value Your Privacy ... Now Take Some Cookies. *Informatik Spektrum* 42 (2018), 345 – 346. https://api.semanticscholar.org/CorpusID:52011984

[14] Per G. Fredriksson and Daniel L. Millimet. 2002. Is There a 'California Effect' in US Environmental Policymaking? *Regional Science and Urban Economics* 32, 6 (2002), 737–764. https://doi.org/10.1016/S0166-0462(01)00096-5

[15] Eric Goldman. 2020. An Introduction to the California Consumer Privacy Act (CCPA). Santa Clara Univ. Legal Studies Research Paper. Available at SSRN: https://ssrn.com/abstract=3211013 or http://dx.doi.org/10.2139/ssrn.3211013.

[16] Cymone Gosnell. 2019. The General Data Protection Regulation: American Compliance Overview and the Future of the American Business. *J. Bus. & Tech.* 15, 1 (2019), 165–187. https://doi.org/10.1137/080734467

[17] Hana Habib, Sarah Pearman, Jiamin Wang, Yixin Zou, Alessandro Acquisti, Lorrie Faith Cranor, Norman Sadeh, and Florian Schaub. 2020. "It's a scavenger hunt": Usability of Websites' Opt-Out and Data Deletion Choices. In *Proceedings of the 2020 CHI Conference on Human Factors in Computing Systems* (<conf-loc>, <city>Honolulu</city>, <state>HI</state>, <country>USA</country>, </conf-loc>) *(CHI '20)*. Association for Computing Machinery, New York, NY, USA, 1–12. https://doi.org/10.1145/3313831.3376511

[18] Hana Habib, Yixin Zou, Aditi Jannu, Neha Sridhar, Chelse Swoopes, Alessandro Acquisti, Lorrie Faith Cranor, Norman Sadeh, and Florian Schaub. 2019. An Empirical Analysis of Data Deletion and Opt-Out Choices on 150 Websites. In *Fifteenth Symposium on Usable Privacy and Security (SOUPS 2019)*. USENIX Association, Santa Clara, CA, 387–406. https://www.usenix.org/conference/soups2019/presentation/habib

[19] Hana Habib, Yixin Zou, Yaxing Yao, Alessandro Acquisti, Lorrie Cranor, Joel Reidenberg, Norman Sadeh, and Florian Schaub. 2021. Toggles, Dollar Signs, and Triangles: How to (In)Effectively Convey Privacy Choices with Icons and Link Texts. In *Proceedings of the 2021 CHI Conference on Human Factors in Computing*

*Systems* (Yokohama, Japan) *(CHI '21)*. Association for Computing Machinery, New York, NY, USA, Article 63, 25 pages. https://doi.org/10.1145/3411764.3445387

[20] S. Human, H. J. Pandit, V. Morel, C. Santos, M. Degeling, A. Rossi, W. Botes, V. Jesus, and I. Kamara. 2022. Data Protection and Consenting Communication Mechanisms: Current Open Proposals and Challenges. In *2022 IEEE European Symposium on Security and Privacy Workshops (EuroS&PW)*. IEEE Computer Society, Los Alamitos, CA, USA, 231–239. https://doi.org/10.1109/EuroSPW55150.2022.00029

[21] Jane Im, Ruiyi Wang, Weikun Lyu, Nick Cook, Hana Habib, Lorrie Faith Cranor, Nikola Banovic, and Florian Schaub. 2023. Less is Not More: Improving Findability and Actionability of Privacy Controls for Online Behavioral Advertising. In *Proceedings of the 2023 CHI Conference on Human Factors in Computing Systems* (<conf-loc>, <city>Hamburg</city>, <country>Germany</country>, </conf-loc>) *(CHI '23)*. Association for Computing Machinery, New York, NY, USA, Article 661, 33 pages. https://doi.org/10.1145/3544548.3580773

[22] Meg Leta Jones and Margot E Kaminski. 2020. An American's Guide to the GDPR. *Denv. L. Rev.* 98 (2020), 93. https://ssrn.com/abstract=3620198

[23] Cameron F. Kerry. 2023. Will California be the death of National Privacy Legislation? https://www.brookings.edu/articles/will-california-be-the-death-of-national-privacy-legislation/.

[24] Joan M Kiel and Laura M Knoblauch. 2010. HIPAA and FERPA: Competing or Collaborating? *Journal of Allied Health* 39, 4 (2010), 161E–165E. https://www.ingentaconnect.com/content/asahp/jah/2010/00000039/00000004/art00015

[25] Michael Kretschmer, Jan Pennekamp, and Klaus Wehrle. 2021. Cookie Banners and privacy policies: Measuring the Impact of the GDPR on the Web. *ACM Transactions on the Web (TWEB)* 15, 4 (2021), 1–42.

[26] McKenzie L Kuhn. 2018. 147 Million Social Security Numbers for Sale: Developing Data Protection Legislation after Mass Cybersecurity Breaches. *Iowa L. Rev.* 104 (2018), 417. https://ilr.law.uiowa.edu/print/volume-103-issue-6/147-million-social-security-numbers-for-sale-developing-data-protection-legislation-after-mass-cybersecurity-breaches

[27] Filippo Lancieri. 2022. Narrowing Data Protection's Enforcement Gap. *Me. L. Rev.* 74 (2022), 15.

[28] Vincent Lefrere, Logan Warberg, Cristobal Cheyre, Veronica Marotta, and Alessandro Acquisti. 2020. *The impact of the GDPR on content providers*. Post-Print hal-03111801. HAL. https://ideas.repec.org/p/hal/journl/hal-03111801.html

[29] Not Listed. 2023. Attorney General Bonta, Governor Newsom and CPPA - Letter Opposing Federal Privacy Preemption. https://cppa.ca.gov/pdf/adppa_letter.pdf.

[30] Not Listed. 2023. Brave web browser. https://brave.com/.

[31] Not Listed. 2023. The California Consumer Privacy Act. https://theccpa.org/. [Online; accessed 28-November-2023].

[32] Not Listed. 2023. California Consumer Privacy Act (CCPA). https://oag.ca.gov/privacy/ccpa. [Online; accessed 28-November-2023].

[33] Not Listed. 2023. California Privacy Protection Agency - california consumer privacy act. https://cppa.ca.gov/regulations/pdf/cppa_act.pdf. [Online; accessed 28-November-2023].

[34] Not Listed. 2023. Colorado Privacy Act (CPA). https://coag.gov/resources/colorado-privacy-act/. [Online; accessed 28-November-2023].

[35] Not Listed. 2023. DuckDuckGo web browser. https://duckduckgo.com/.

[36] Not Listed. 2023. Multiple US states advance privacy legislation; Illinois House fails to Pass Consumer Privacy Act. https://iapp.org/news/a/multiple-us-states-advance-privacy-legislation-illinois-house-fails-to-pass-consumer-privacy-act/.

[37] Not Listed. 2023. Powerful API Stack For Data-Driven Marketers. http://www.dataforseo.com/.

[38] Not Listed. 2023. State of California - Department of Justice - Office of the Attorney General. https://oag.ca.gov/privacy/ccpa.

[39] Not Listed. 2023. Utah Consumer Privacy Act. https://le.utah.gov/~2022/bills/static/SB0227.html. [Online; accessed 28-November-2023].

[40] Not Listed. 2023. Venture Capital, private equity and M&A database | Pitchbook. https://pitchbook.com/.

[41] Not Listed. 2023. The Virginia Consumer Data Protection Act - oag.state.va.us. https://www.oag.state.va.us/consumer-protection/files/tips-and-info/Virginia-Consumer-Data-Protection-Act-Summary-2-2-23.pdf.

[42] Not Listed. 2023. Zoominfo. https://www.zoominfo.com/.

[43] Maureen Mahoney. 2020. CCPA: Are Consumers' Digital Rights Protected? https://advocacy.consumerreports.org/wp-content/uploads/2020/09/CR_CCPA-Are-Consumers-Digital-Rights-Protected_092020_vf.pdf. Accessed: 2023-11-08.

[44] Trung Tin Nguyen, Michael Backes, and Ben Stock. 2022. Freely Given Consent? Studying Consent Notice of Third-Party Tracking and Its Violations of GDPR in Android Apps. In *Proceedings of the 2022 ACM SIGSAC Conference on Computer and Communications Security* (Los Angeles, CA, USA) *(CCS '22)*. Association for Computing Machinery, New York, NY, USA, 2369–2383. https://doi.org/10.1145/3548606.3560564

[45] Midas Nouwens, Ilaria Liccardi, Michael Veale, David Karger, and Lalana Kagal. 2020. Dark Patterns after the GDPR: Scraping Consent Pop-ups and Demonstrating their Influence. In *Proceedings of the 2020 CHI Conference on Human Factors in Computing Systems* (<conf-loc>, <city>Honolulu</city>, <state>HI</state>, <country>USA</country>, </conf-loc>) *(CHI '20)*. Association for Computing Machinery, New York, NY, USA, 1–13. https://doi.org/10.1145/3313831.3376321

[46] Sean O'Connor, Ryan Nurwono, Aden Siebel, and Eleanor Birrell. 2021. (Un)clear and (In)conspicuous: The Right to Opt-out of Sale under CCPA. In *Proceedings of the 20th Workshop on Workshop on Privacy in the Electronic Society* (Virtual Event, Republic of Korea) *(WPES '21)*. Association for Computing Machinery, New York, NY, USA, 59–72. https://doi.org/10.1145/3463676.3485598

[47] Nicholas F Palmieri III. 2020. Who Should Regulate Data: An Analysis of the California Consumer Privacy Act and Its Effects on Nationwide Data Protection Laws. *Hastings Sci. & Tech. LJ* 11 (2020), 37. https://papers.ssrn.com/sol3/papers.cfm?abstract_id=3521507

[48] Stuart L Pardau. 2018. The California Privacy Act: Towards a European-Style Privacy Regime in the United States. *J. TECH. L. & POL'Y* 23 (2018), 68.

[49] Irwin Reyes, Primal Wijesekera, Joel Reardon, Amit Elazari Bar On, Abbas Razaghpanah, Narseo Vallina-Rodriguez, and Serge Egelman. 2018. "Won't Somebody Think of the Children?" Examining COPPA Compliance at Scale. *Proceedings on Privacy Enhancing Technologies* 2018 (2018), 63 – 83. https://api.semanticscholar.org/CorpusID:4935390

[50] Paul M Schwartz. 2013. The EU-U.S. Privacy Collision: A Turn to Institutions and Procedures. *Harvard Law Review* 126 (2013), 1966–2009.

[51] Paul M Schwartz and Karl-Nikolaus Pfeifer. 2017. Transatlantic Data Privacy Law. *Georgetown Law Journal* 106 (2017), 115–179.

[52] A. Millar Sheila and P. Marshall Tracy. 2023. The State of the State Privacy Laws: A Comparison | Keller and Heckman. https://www.khlaw.com/insights/state-state-privacy-laws-comparison. [Online; accessed 28-November-2023].

[53] Aden Siebel and Eleanor Birrell. 2022. The Impact of Visibility on the Right to Opt-out of Sale under CCPA. https://arxiv.org/abs/2206.10545. arXiv:2206.10545 [cs.CY]

[54] Daniel J. Solove. 2013. HIPAA Turns 10: Analyzing the Past, Present, and Future Impact. *Journal of AHIMA* 84 (April 2013), 22–28. https://papers.ssrn.com/sol3/papers.cfm?abstract_id=2245022 GWU Legal Studies Research Paper No. 2013-75, GWU Law School Public Law Research Paper No. 2013-75, Available at SSRN: https://ssrn.com/abstract=2245022.

[55] Daniel J Solove. 2017. The Limitations of Privacy RightsThe Limitations of Privacy R. *Notre Dame Law Review* 98 (2017), 975–1036.

[56] P. Augustinos Theodore and R. Cox Alexander. 2023. U.S. State Privacy Laws in 2023: California, Colorado, Connecticut, Utah and Virginia. https://www.lockelord.com/newsandevents/publications/2022/12/us-state-privacy-laws-2023. [Online; accessed 28-November-2023].

[57] Maggie Van Nortwick and Christo Wilson. 2022. Setting the Bar Low: Are Websites Complying With the Minimum Requirements of the CCPA? *Proceedings on Privacy Enhancing Technologies* 2022 (2022), 608––628.

[58] Ari Ezra Waldman. 2017. Privacy's Rights Trap. *Northwestern University Law Review* 117 (2017), 88–106.

[59] Richmond Y. Wong and Deirdre K. Mulligan. 2019. Bringing Design to the Privacy Table: Broadening "Design" in "Privacy by Design" Through the Lens of HCI. In *Proceedings of the 2019 CHI Conference on Human Factors in Computing Systems* (Glasgow, Scotland Uk) *(CHI '19)*. Association for Computing Machinery, New York, NY, USA, 1–17. https://doi.org/10.1145/3290605.3300492

[60] Sebastian Zimmeck, Oliver Wang, Kuba Alicki, Jocelyn Wang, and Sophie Eng. 2023. Usability and Enforceability of Global Privacy Control. *Proceedings on Privacy Enhancing Technologies* 2 (2023), 1–17. https://www.petsymposium.org/2023/files/papers/issue2/popets-2023-0052.pdf

# A APPENDIX

**Table 7: Comparison of state specific privacy laws' jurisdiction applicability and jurisdiction details around the sale of personal information for CA, VA, CO, and UT, updated by July 31, 2023.**

| Category | Criteria | The CCPA (CA) | The VCDPA (VA) | The CPA (CO) | The UCPA (UT) |
|---|---|---|---|---|---|
| Jurisdiction applicability | Type of product or service | Conducts business in the state or produces a product or service that is targeted to consumers who are residents of the state | | | |
| | Business type | For-profit | For-profit | Include non-profit | Not mention |
| | Business characteristics | **Either:** (1) At least $25 million in annual revenue (2) Handle personal data of 100,000 or more annually (3) Earn at least 50% of annual revenue from selling personal data | **Either:** (1) Handle personal data of 100,000 or more annually (2) Handle personal data of 25,000 or more annually and earn at least 50% of annual revenue from selling personal data | **Either:** (1) Handle personal data of 100,000 or more annually (2) Derives revenue or discounts from selling personal data of 25,000 or more | At least $25 million in annual revenue **AND Either:** (1) Handle personal data of 100,000 or more annually (2) Handle personal data of 25,000 or more annually and earn at least 50% of annual revenue from selling personal data |
| Jurisdiction details | Opt-out rights | Opt out of selling or sharing of personal information | Opt out of targeted advertising, sales of personal information or further profiling | | |
| | Opt-out methods | Using specific opt-out methods such as the conspicuous opt-out link on website's main page or the friction-less opt-out preference signals | Submit request to the controller via methods prescribed by the controller | | |